\documentclass[prd,twocolumn,showpacs,floatfix,amsmath,nofootinbib,amssymb,floatfix]{revtex4}
\usepackage{graphicx,color,dcolumn,booktabs,bm,multirow}
\usepackage{longtable,lscape}
\usepackage{txfonts}
\usepackage{overpic}
\usepackage{amssymb}
\usepackage{indentfirst}
\usepackage{feynmf}   
\usepackage{slashed}  
\usepackage{cases}
\usepackage{color}
\usepackage{multirow}
\usepackage{epstopdf}
\usepackage{graphicx,color,dcolumn,booktabs,bm}
\usepackage{epstopdf}

\usepackage[colorlinks,
            citecolor=green,
            anchorcolor=red,
            menucolor=red,
            linkcolor=red,
            filecolor=red,
            runcolor=red,
            urlcolor=blue,
            frenchlinks=red]{hyperref}

\begin{document}

\title{Reconciling the X(2240) with the Y(2175)}
\author{Dian-Yong Chen$^{1}$\footnote{Corresponding author}}\email{chendy@seu.edu.cn}
\author{Jing Liu$^1$}
\author{Jun He$^2$}
\affiliation{
$^1$ School of Physics, Southeast University, Nanjing 210094, People's Republic of China\\
$^2$ Department of Physics and Institute of Theoretical Physics, Nanjing Normal University, Nanjing 210097, People's Republic of China
}

\date{\today}

\begin{abstract}
In the present work, we reanalyzed the cross sections for $e^+ e^- \to K^+K^-$, where a new structure $X(2240)$ was reported by BES III Collaboration. By including the interference between the direct coupling and vector meson intermediate processes, we find the mass and width of $X(2240)$ are $2197.4\pm 4.4$ MeV and $75.6\pm 7.2 $ MeV, respectively, which are well consistent with the PDG average values of the resonance parameters for $Y(2175)$, thus, we conclude that the  $X(2240)$ should be the same state as the $Y(2175)$.
\end{abstract}
\maketitle

\section{Introduction}\label{sec1}
In the past decades, an increasing number of new hadron states have been observed in charm and bottom sectors (see Refs~\cite{Klempt:2007cp,Brambilla:2010cs,Chen:2016qju,Lebed:2016hpi,  Guo:2017jvc,Esposito:2016noz,Ali:2017jda, Liu:2019zoy,Brambilla:2019esw,Dong:2017gaw} for recent reviews of experimental and theoretical status). These experimental observations stimulate theorists great interests in hunting QCD exotic states in the forest of heavy-quarkonium-like states. However, in the light flavor sector, the researches of the new hadron states in both experimental and theoretical sides are not as prosperous as those in heavy flavor sectors. As one of the few interesting new hadron states in light sector, the strangeonium-like state $Y(2175)$ was first reported by BaBar Collaboration in $e^+ e^- \to \gamma_{\mathrm{ISR}}\phi f_0 (980)$ in 2006~\cite{Aubert:2006bu}. And then it was confirmed by BESII Collaboration in the $J/\psi \to \eta \phi f_0(980)$ \cite{Ablikim:2007ab} and by Belle Collaboration in the $e^+ e^-\to \phi  f_0(980)$ and $\phi  \pi^+ \pi^-$ processes~\cite{Shen:2009zze}. 

\begin{figure}[t]
\centering
\scalebox{0.8}{\includegraphics{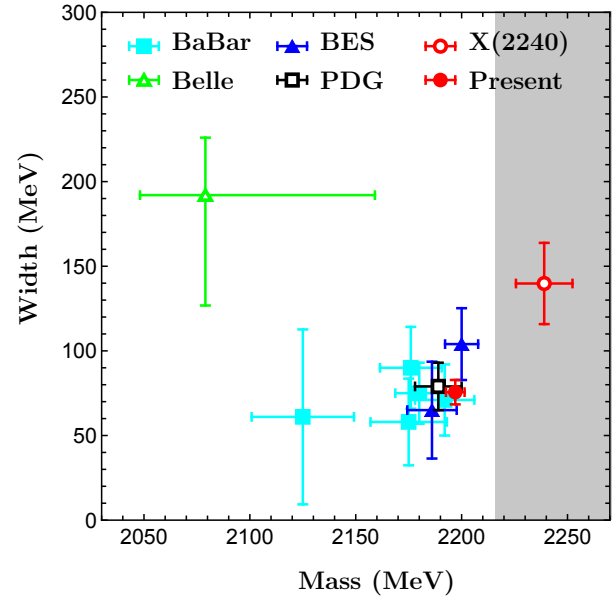}}	
\caption{(Color online.) The resonance parameters of $Y(2175)$ reported by BaBar~\cite{Lees:2011zi, Aubert:2007ym, Aubert:2007ur, Aubert:2006bu}, Belle~\cite{Shen:2009zze} and BES~\cite{Ablikim:2007ab, Ablikim:2014pfc} Collaborations, as well as the PDG average~\cite{Tanabashi:2018oca}. For comparison, we also present the resonance parameters of $X(2240)$ reported by BES III Collaboration~\cite{Ablikim:2018iyx}. \label{Fig:Mass}}
\end{figure}

Since its discovery, $Y(2175)$ has been searched in $e^+ e^- \to \gamma_{\mathrm{ISR}} \phi  \pi^+ \pi^-$, $e^+ e^- \to \gamma_{\mathrm{ISR}} \phi  \eta $, $e^+e^- \to \eta  \phi  f_0(980)$, $e^+ e^- \to \gamma_{\mathrm{ISR}} \phi  f_0(980)$, $e^+ e^- \to \gamma_{\mathrm{ISR}} K^+ K^- f_0(980)$, $J/\psi \to \eta  \phi  f_0(980)$ $e^{+}e^{-} \to K^{+}K^{-}K^{+}K^{-} $,$e^{+}e^{-} \to K^{+}K^{-}\pi^{0} \pi^{0} $ and  $e^{+}e^{-} \to \phi \eta^\prime $ by BaBar~\cite{Lees:2011zi, Aubert:2007ym, Aubert:2007ur, Aubert:2006bu}, Belle~\cite{Shen:2009zze} and BES~\cite{Ablikim:2007ab, Ablikim:2014pfc, Ablikim:2019tpp, Ablikim:2020pgw, Ablikim:2020coo} Collaborations. The resonance parameters reported by different collaborations are presented in Fig.~\ref{Fig:Mass}. From the figure, one can find the reported mass of $Y(2175)$ is mostly located in the vicinity of 2180 MeV, while its width is about 80 MeV. The PDG average of its resonance parameters are~\cite{Tanabashi:2018oca}, 
\begin{eqnarray}
	m_{Y(2175)}&=& (2188 \pm 10 )\ \mathrm{MeV}, \nonumber\\
    \Gamma_{Y(2175)} &=& (83 \pm 12)\ \mathrm{MeV},	
    \label{Eq:PDG-Y2175}
\end{eqnarray}
respectively.

At the early stage, the strangeonium-like state, $Y(2175)$, was only observed in the hidden-strange process, which is similar to the case of $Y(4260)$ in $ \pi^+ \pi^- J/\psi$~\cite{Aubert:2005rm}, $Y(4360)$ in  $\pi^+\pi^- \psi(2S)$~\cite{Aubert:2007zz} and $Y_b(10890)$ in $\Upsilon(nS) \pi^+ \pi^-,\ (n=1,2)$ processes~\cite{Abe:2007tk}. The lack of the observation in the open-strange channels makes this state particular interesting. In Ref.~\cite{Wang:2006ri}, by using QCD sum rule approach and take more phenomenological analysis, the mass of $Y(2175)$ can be reproduced in a tetraquark frame. As indicated in Ref.~\cite{Drenska:2008gr}, the strangeonium-like state $Y(2175)$ can also be assigned as an excited tetraquark state with spin 1 and the estimation in the flux-tube model supported that $Y(2175)$ is a tetraquark state with  diquark-anti-diquark configuration~\cite{Deng:2010zzd}. Considering the particular production processes of $Y(2175)$, a hybrid interpretation was proposed in Refs.~\cite{Ding:2006ya, Ding:2007pc}, and the estimations in both flux-tube model and constituent gluon model supported such an interpretation. But the non-perturbative Lattice estimation indicated that $Y(2175)$ could not be a $s\bar{s} g$ hybrid due to its decay behavior~\cite{Dudek:2011bn}. The recent estimations with QCD Gaussian sum rule also find that $Y(2175)$ can hardly be assigned as a hybrid due to the small relative strength~\cite{Ho:2019org}. Moreover, in the vicinity of $Y(2175)$, there are some thresholds of a pair of hadrons, such as $\Lambda \bar{\Lambda}$, $\phi f_0(980)$, thus,  in Ref.~\cite{Klempt:2007cp}, $Y(2175)$ was interpreted as a $\Lambda \bar{\Lambda}$ baryonium and in Ref.~\cite{MartinezTorres:2008gy}, $Y(2175)$ was considered as a resonance of $\phi K \bar{K}$. Besides these exotic interpretations, $Y(2175)$ was also tried to be categorized as a conventional strangeonium, such as $3^3S_1$~\cite{Barnes:2002mu} or $2^3D_1$~\cite{Ding:2006ya,Wang:2012wa} states.

Among these interpretations, the open strange decays are important criteria, thus searching the signal of $Y(2175)$ experimentally in the open strange channels are crucial. In 2018, the BES III Collaboration measured the cross sections for $e^+ e^- \to K^+ K^-$ at $\sqrt{s}=  2.0 - 3.08 $ GeV with the best precision achieved so far, no signal of $Y(2175)$ was found but a new structure, named $X(2240)$, was reported with the resonance parameters to be \cite{Ablikim:2018iyx},
\begin{eqnarray}
m_{X(2240)} &=& 2239.2 \pm 7.1 \pm 11.3 \ \mathrm{MeV},\nonumber\\
\Gamma_{X(2240)} &=& 139.8 \pm 12.3 \pm 20.6 \ \mathrm{MeV},
\end{eqnarray}
respectively. In Fig.~\ref{Fig:Mass}, we also present the resonance parameters of $X(2240)$ for comparison, which differs from PDG average values of $Y(2175)$ resonance parameters by more than $3\sigma$ in mass and more than $2 \sigma$ in width. After the observation of $X(2400)$, it has been investigate in different scenario, such as tetraquark~\cite{Azizi:2019ecm, Lu:2019ira}, $\Lambda\bar{\Lambda}$ baryonium~\cite{Zhu:2019ibc}, $\omega(3^3D_1)$ state~\cite{Wang:2019jch}. 

If carefully checking the observation of $Y(2175)$ and $X(2240)$, one can find that all the two-body discovery channels of $Y(2175)$ are neutral, such as $\phi \eta$, $\phi f_0(980)$. Even the three-body channel of $Y(2175)$, such as $\phi \pi^+ \pi^-$ and $K^+ K^- f_0(980)$, the charged $\pi^+ \pi^-$ and $K^+ K^-$ can form a neutral bloc. However, as for $X(2240)$, it was discovered in a charged $K^+ K^-$ channel, which indicates that the virtual photon from the $e^+ e^-$ annihilation can couple with the $K^+K^-$ pair directly, or the virtual photon couples to the vector resonance and the vector resonance decay into $K^+K^-$. These two kinds of mechanisms working together in the $e^+ e^- \to K^+ K^-$ process and their interferences can change the peak of the resonance. Such phenomena have been applied in the investigations of the cross sections for some processes involved charmonium-like states~\cite{Chen:2010nv, Chen:2011kc, Chen:2015bft, Chen:2017uof}. Moreover, besides $Y(2175)$, there is another state, $\rho(2150)$ in the vicinity of 2.2 GeV. Its mass was reported from $1990 \pm 80$ MeV~\cite{Aubert:2007ef} to $2254 \pm 22$ MeV~\cite{Lees:2012cj} , but almost all the measurement indicate that $\rho(2150)$ are much broader with the width to be several hundreds MeV~\cite{Aubert:2007ef,Tanabashi:2018oca}. The experimental data from BES III Collaboration indicate that there is one peak in the vicinity of 2.2 GeV, thus in the present work, we tried to reproduce the cross sections for $e^+e^- \to K^+K^-$ with the interference between the direct and the resonance contributions. In the fit, we treat the resonance parameters as free parameters, which can be determined by fitting the cross sections for $e^+e^-\to K^+K^-$. By comparing the fitted resonance parameters with those of $Y(2175)$ and $\rho(2150)$, one can conclude the source of the peak near 2.2 GeV. 

This work is organized as follows. After introduction, we present the concrete formula of the interference mechanism in the following section. In Section~\ref{Sec:Num}, we present our fit results and the related discussions and the last section is devoted to a short summary.

\section{Interference Picture of $e^+e^- \to K^+ K^-$}
\label{Sec:Int}

\begin{figure}[htb]
\centering
\begin{tabular}{cc}
\scalebox{0.4}{\includegraphics{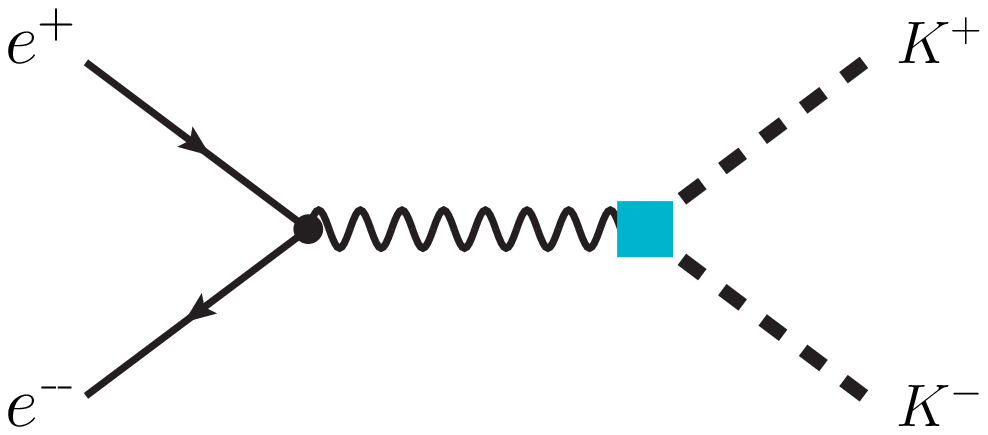}}& 
\scalebox{0.4}{\includegraphics{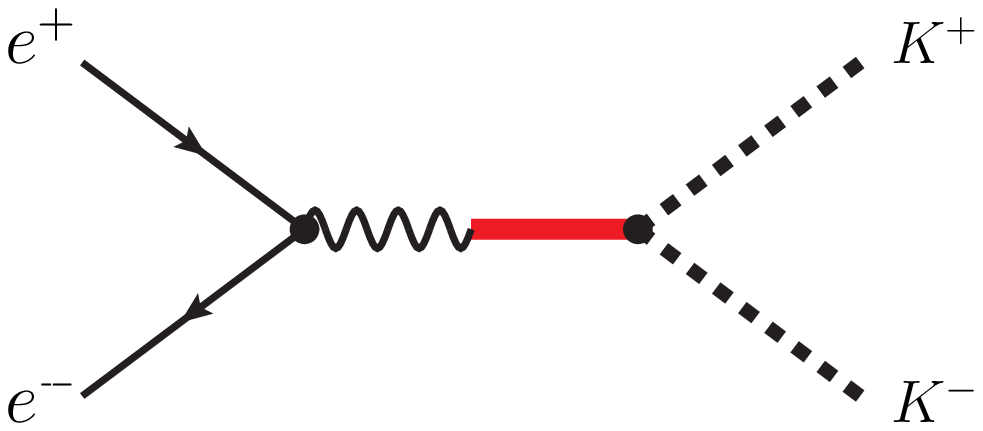}}\\
(a)& (b)
\end{tabular}	
\caption{(Color online.) The mechanisms working in $e^+ e^- \to K^+ K^-$ process. Diagram (a) is the direct coupling process, where the cyan diamond indicate the form factor of kaon. Diagram (b) is the resonance intermediate process, where the red line indicate the intermediate vector meson. \label{Fig:Mech}}
\end{figure}

As indicated in the introduction, there exist two kinds of mechanisms working in $e^+e^-\to K^+ K^-$ process as shown in Fig.~\ref{Fig:Fit}, which are the direct coupling process as shown in diagram (a) and the vector resonance intermediate process as shown in digram (b). In the present work, we estimate these diagrams in hadron level, and the interaction vertexes are depicted by effective Lagrangians. The involved effective Lagrangians are~\cite{Chen:2011cj,Bauer:1975bw, Bauer:1975bv, Bauer:1977iq},
\begin{eqnarray}
\mathcal{L}_{KK\gamma} &=& ie A^\mu (\bar{K} \partial_\mu K-\partial_\mu \bar{K} K)\nonumber\\
\mathcal{L}_{VKK} &=& i g_{VKK} V^\mu (\bar{K} \partial_\mu K-\partial_\mu \bar{K} K), \nonumber\\
\mathcal{L}_{\gamma V} &=&-e \frac{m_V^2}{f_V} V^\mu A_\mu. \label{Eq:Lag}
\end{eqnarray}
With the above effective Lagrangians, we can get the amplitudes corresponding to diagrams in Fig.~\ref{Fig:Mech}, which are, 
\begin{eqnarray}
\mathcal{M}_{\mathrm{Dir}} &=& \Big[\bar{v}(p_2,m_e) (ie \gamma^\mu ) u(p_2,m_e)\Big] \frac{-g_{\mu \nu}}{q^2} \Big[ie (p_4^\nu -p_3^\nu) F_K(q^2)\Big], \nonumber\\
\mathcal{M}_{\mathrm{Res}} &=& \Big[\bar{v}(p_2,m_e) (ie \gamma_\mu ) u(p_2,m_e)\Big] \frac{-g^{\mu \rho}}{q^2}\Big(-e \frac{m_V^2}{f_V} \Big) \nonumber\\
&&\frac{-g_{\rho \nu}+q_\rho q_\nu/m_V^2}{q^2-m_V^2+i m_V \Gamma_V}  \Big[ig_{VKK} (p_4^\nu -p_3^\nu) \Big],	
\end{eqnarray} 
where $q=p_1+p_2$, $F_K$ is the time-like form factor of charged kaon. The total amplitude of $e^+e^- \to K^+ K^-$ is,
\begin{eqnarray}
\mathcal{M}_{\mathrm{Tot}}&=&\mathcal{M}_{\mathrm{Dir}}+e^{i\phi} \mathcal{M}_{\mathrm{Res}}\nonumber\\
&=&\Big[\bar{v}(p_2,m_e) (ie \gamma^\mu ) u(p_2,m_e)\Big] \frac{1}{q^2} \Big[ie (p_4^\nu -p_3^\nu) \Big]\nonumber\\
&&\times \Big[-F_K(q^2)g_{\mu \nu} -e^{i\phi} g_V \frac{-g_{\mu \nu }+q_\mu q_\rho/m_V^2}{q^2-m_V^2+im_V\Gamma_V}  \Big]
\end{eqnarray}
where $\phi$ is the phase angle between two amplitudes and $g_V= g_{V KK} m_V^2/f_V$.  With the above amplitude, we can get the cross sections for $e^+e^- \to K^+ K^-$. 

In the above amplitude, we can treat $\phi$, $g_V$, $m_V$ and $\Gamma_V$ as free parameters, which can be determined by fitting the cross sections for $e^+ e^-\to K^+ K^-$. Besides these parameters, the form factor of charge kaon is not determined. In principle, the form factor in the time-like region is a complex function~\cite{Yang:2019mzq, Chen:2008hka}. Generally, the argument of the form factor changes slowly when $q^2$ is far away from the threshold. So we can suppose that the argument of the form factor is approximately to be a constant in the considered center-of-mass energy. Thus, the argument of the form factor can be absorbed by the phase angle between two amplitudes and the form factor is treated as a real function of $s$. In the present work, we assume the form factor in the form,
\begin{equation}
	F_K(s)=a s^b e^{-cs}
\end{equation} 
where $a$, $b$, $c$ are considered as free parameters.

\begin{table*}
\caption{The parameter values obtained by fitting the cross sections for $e^+ e^-\to K^+ K^-$. The parameter $a$ is in unit of $\mathrm{GeV}^{-2b}$, which makes the form factor $F_K$ is dimensionless.\label{Tab:Para}}
\begin{tabular}{p{2cm}<{\centering} p{5cm}<{\centering} p{2cm}<{\centering} p{5cm}<{\centering}}
\toprule[1pt]
Parameter & Value &	Parameter & Value\\
\midrule[1pt]
$g_V$     & $(1.33\pm 0.13)\times 10^{-2}\ \mathrm{GeV}^2 $    & $a$ &$2.49 \pm 0.06$  \\
$\phi$    & $(3.06 \pm 0.11)$\ rad   & $b$ & $-2.79 \pm 0.07$ \\
$m_V$     & $(2197.4 \pm 4.4)$\ MeV    & $c$ & $(2.28 \pm 0.71) \times 10^{-2}\ \mathrm{GeV}^{-2}$ \\
$\Gamma_V$& $(75.6 \pm 7.2)$\ MeV\\
\bottomrule[1pt]
\end{tabular}	
\end{table*}

\section{Numerical Results}
\label{Sec:Num}

\subsection{The cross sections for $e^+e^- \to K^+ K^-$}
\begin{figure}[htb]
\centering
\scalebox{0.8}{\includegraphics{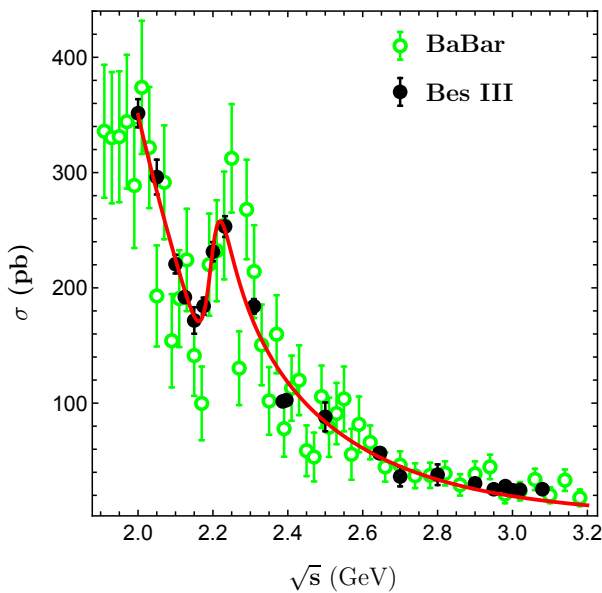}}	
\caption{(Color online.) The fitted cross sections for $e^+ e^- \to K^+ K^-$ depending on the center-of-mass energy. For comparison, we also present the measured data from BES III~\cite{Ablikim:2018iyx} and BaBar Collaboration~\cite{Lees:2013gzt}. \label{Fig:Fit}}
\end{figure}

\begin{figure}[htb]
\centering
\scalebox{0.8}{\includegraphics{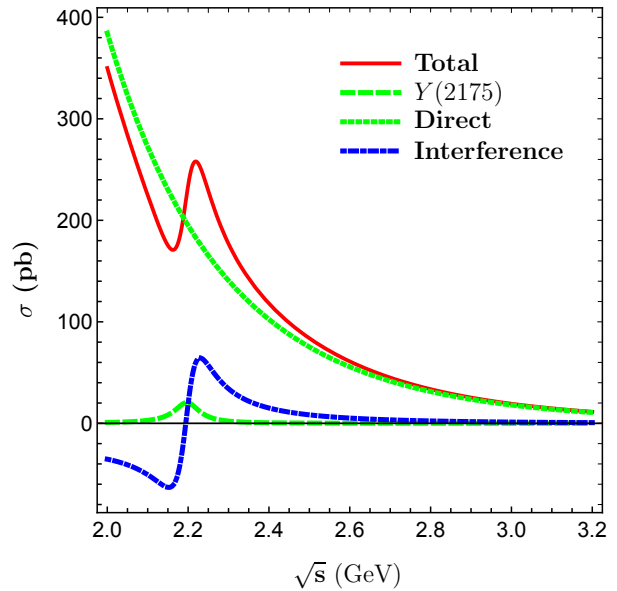}}	
\caption{(Color online.) The individual contributions to the cross sections for $e^+e^-\to K^+ K^-$. The green dashed and dotted curves indicate the contributions from $Y(2175)$ intermediate process and direct coupling process, respectively. The blue dot-dashed curve is the interference between two different contributions. \label{Fig:Ind}}
\end{figure}

With the above preparation, we can fit the cross sections for $e^+e^- \to K^+ K^-$ reported by BES III Collaboration~\cite{Ablikim:2018iyx}, where seven free parameters, $a$, $b$, $c$, $\phi$, $g_V$, $m_V$ and $\Gamma_V$ are determined. It should be noticed that the experimental data from the BES III Collaboration have very small uncertainties ~\cite{Ablikim:2018iyx}, but the continuity of the data is not as good as its precision. In particular, there are only two almost degenerated data from 2.31 to 2.5 GeV, which can not reflect more detail of the cross sections in this energy range. Thus, in the present work, we set the error to be 10 pb for every data. With this assumption, we fit the cross sections for $e^+e^- \to K^+ K^-$ reported by BES III Collaboration~\cite{Ablikim:2018iyx}. The determined parameter values are listed in Table~\ref{Tab:Para}. The errors on the parameters are determined by the contour in the parameter space defined by $\chi^2(\theta_i)=\chi_{\mathrm{min}}^2+1$ with $\theta_i$ to be the undetermined parameters.  With the center values of the fitted parameters, the $\chi^2/\mathrm{n.d.f}$ is estimated to be $16.9/15$. From the present fit, we find the resonance parameters of the vector resonance are,
\begin{eqnarray}
	m_V&=&(2197.4 \pm 4.4)\ \mathrm{MeV},\nonumber\\
	\Gamma_V&=&(75.6 \pm 7.2)\ \mathrm{MeV},
\end{eqnarray} 
respectively, which are well consistent with the PDG average of resonance parameters for $Y(2175)$. In Fig.~\ref{Fig:Mass}, we also present the resonance parameters of this vector state obtained from the present fit. One can find the resonance parameters of this vector state are also consistent with those of $Y(2175)$ reported from BaBar and BES Collaborations \cite{Lees:2011zi, Aubert:2007ym, Aubert:2007ur, Aubert:2006bu, Ablikim:2007ab, Ablikim:2014pfc}. Thus, we conclude that the $X(2240)$ reported by BES III Collaboration should be the same state as the $Y(2175)$.

From the effective Lagragians listed in Eq.~(\ref{Eq:Lag}), one can get the dilepton and $K^+ K^-$ decay widths of $Y(2175)$, which are, 
\begin{eqnarray}
\Gamma_{e^+e^-}&=&\frac{e^4 m_Y}{12 \pi f_Y^2}\nonumber,\\
\Gamma_{K^+K^-} &=& \frac{g_{VKK}^2 (m_Y^2-4m_K^2)^{3/2}}{48 \pi m_Y^2},
\end{eqnarray}
respectively. Thus, one can get the product of two widths, which is,
\begin{eqnarray}
\Gamma_{e^+e^-}\times \Gamma_{K^+K^-} =	\frac{e^4 g_Y^2 (m_Y^2-4m_K^2)^{3/2}}{576 \pi^2 m_Y^5}.\label{Eq:GeeG}
\end{eqnarray}
With this formula and the fitting data, one can get $\Gamma_{e^+e^-} \mathcal{B}(Y\to K^+K^-)$, which is listed in Table~\ref{Tab:GeeB}. As a comparison, we also collect the measured data for other processes. From the table, one can find the branching ratios of $Y(2175)\to K^+K^-$ are at least several times smaller than those of $\phi \eta$, $\phi f_0(980)$ and $\phi \pi^+\pi^-$.

\begin{table}[t]
\caption{A comparison of $\Gamma_{e^+e^-} \mathcal{B}(Y\to f)$ with $f=(K^+ K^-,\ \phi f_0(980),\ \phi \eta,\ \phi \pi^+\pi^-)$. The one for $K^+K^-$ is obtained by Eq.~(\ref{Eq:GeeG}) with the center values of the fitting parameters.\label{Tab:GeeB}}
	\begin{tabular}{p{5cm}<{\centering} p{3cm}<{\centering}}
\toprule[1pt]
Process & Value (eV)\\
\midrule[1pt]
$\Gamma_{e^+e^-}\mathcal{B}(Y\to \phi f_0(980))$ &	
$2.5\pm 0.8\pm 0.4$~\cite{Aubert:2007ur}\\
&	
$2.3\pm 0.3\pm 0.3$~\cite{Lees:2011zi}\\
$\Gamma_{e^+e^-}\mathcal{B}(Y\to \phi \eta)$ &	
$1.7\pm 0.7\pm 1.3$~\cite{Aubert:2007ym}\\
$\Gamma_{e^+e^-}\mathcal{B}(Y\to \phi \pi^+ \pi^-)$ &	
$2.90$\footnote{In Ref.~\cite{Lees:2011zi}, BaBar Collaboration reported the cross sections for $e^+ e^-\to \phi \pi^+ \pi^-$ and the fitted mass and width of $Y(2175)$ in this channel is $2176 \pm 14 \pm 4$ and $90 \pm 22\pm 10$ MeV, respectively. The cross sections resulted from $Y(2175)$ is fitted to be $0.082 \pm 0.024 \pm 0.010$ nb. The product of the dilepton width and branching fraction to $\phi \pi^+\pi^-$ is $\Gamma_{e^+ e^-} \mathcal{B}(\phi \pi^+ \pi^-)=\Gamma_Y \sigma m_Y^2/(12 \pi C_0))$ with $C_0=0.389\ \mathrm{mb (GeV)}^2$. With the center values of the mass, width and cross sections, we roughly estimate the product $\Gamma_{e^+ e^-} \mathcal{B}(\phi \pi^+\pi^-)= 2.90$ eV.}~\cite{Lees:2011zi}\\
& $18.1\pm 1.8$\footnote{In Ref.\cite{Shen:2009mr}, the authors performed a combined fit to BaBar and Belle data on $e^+e^-\to \phi \pi^+ \pi^-$. }~\cite{Shen:2009mr} \\
\midrule[1pt]
$\Gamma_{e^+e^-}\mathcal{B}(Y\to K^+K^-)$ &	
$0.51$\\
\bottomrule[1pt]
	\end{tabular}
\end{table}

With the central values of fitted parameters, we can obtain the cross sections for $e^+e^- \to K^+K^-$, which is shown as the red curve in Fig.~\ref{Fig:Fit}. The fitted curve can well reproduce the experimental data of the cross sections for $e^+ e^- \to K^+ K^-$. However, the fitted curve is a bit larger than the experimental data around 2.4 GeV. As we indicated at the beginning of this section, there are no enough experimental data around this area, we can not get the detail of the lineshape in the vicinity of 2.4 GeV, especially, some measurements have indicated that there may be a new state around 2.45 GeV~\cite{Aubert:2007ur, Ablikim:2007ab, Shen:2009zze, Ablikim:2014pfc, Shen:2009mr}. We expect the BES III Collaboration could provide more precise data around 2.4 GeV, which could help us to identify the existence of new structure in this area.

We present the individual contributions to the cross sections in Fig. \ref{Fig:Ind}. From the figure, one can find the dominant contribution comes from the direct coupling process and the contribution from $Y(2175)$ is small. However, the interference between these two mechanisms is very important to reproduce the experimental data, which shifts the peak of $Y(2175)$ to 2240 MeV.

\subsection{The cross sections for $e^+ e^- \to \pi^+ \pi^-$}

\begin{figure}[htb]
\centering
\scalebox{0.8}{\includegraphics{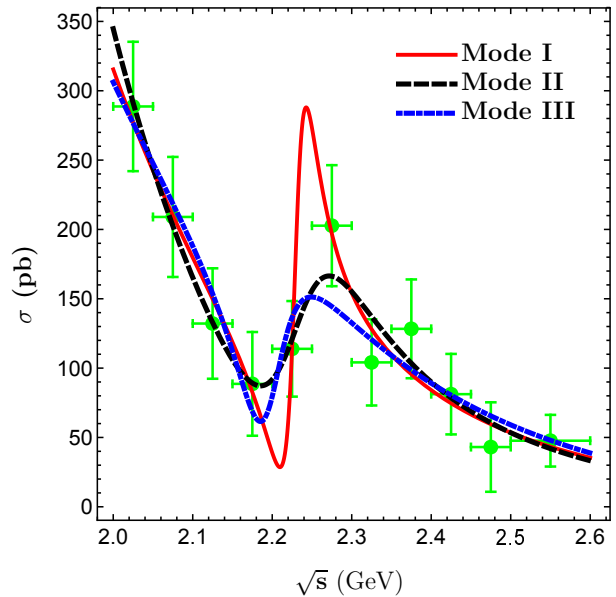}}	
\caption{(Color online.) The fitted cross sections of $e^+ e^- \to \pi^+ \pi^-$ depending on the center-of-mass energy.
 \label{Fig:Fit-pipi}}
\end{figure}

\begin{table*}
\caption{The same as Table \ref{Tab:Para} but for the cross sections for $e^+e^- \to \pi^+ \pi^-$ with different fitting modes.\label{Tab:Para-pipi}}
\begin{tabular}{p{3cm}<{\centering} p{4cm}<{\centering} p{4cm}<{\centering} p{4cm}<{\centering}}
\toprule[1pt]
Parameter & Mode I &	 Mode II & Mode III\\
\midrule[1pt]
$a$ & $(7.71 \pm 2.88) \times 10^{-2}$ & $(1.66\pm 0.96)\times 10^{-1}$ & $(2.11\pm 1.10)\times 10^{-2}$ \\
$b$ & $-2.68 \pm 0.93$ & $-3.89 \pm 1.59$ & $(8.97\pm 20.1) \times 10^{-1}$ \\
$c\ (\mathrm{GeV}^{-2}) $ & $(9.56 \pm 13.7) \times 10^{-2}$ & $(5.54 \pm 1.65) \times 10^{-2}$ & $(3.96 \pm 1.54) \times 10^{-1} $ \\
$g_V\ (\mathrm{GeV}^2)$ & $(4.56\pm 2.73)\times 10^{-4} $ & $(9.18 \pm 2.79) \times 10^{-4}$ & $(4.94\pm 1.97) \times 10^{-4}$ \\  
$\phi$ (rad) & $3.48 \pm 0.68$ & $ 3.42 \pm 0.27$ & $4.20 \pm 0.31$\\
$m_V$ (MeV)  & $2230.1 \pm 8.8$  & 2232 (fixed)   & 2188 (fixed)\\
$\Gamma_V$ (MeV) & $34.3 \pm 72.1$ & 133 (fixed) & 83 (fixed)\\
 \midrule[1pt]
$\chi^2$  & 2.11 & 2.78 & 4.55 \\
\bottomrule[1pt]
\end{tabular}	
\end{table*}

In the above fit, we find the interferences are crucial in understanding the source the structure near 2.2 GeV. Such kind of interferences should also appear in the similar process, such as $e^+ e^- \to \pi^+ \pi^-$. But unfortunately, the cross sections for this process are not as precise as those for $e^+ e^- \to K^+ K^-$. In Refs.~\cite{BABAR:2019oes, Lees:2012cj}, the BaBar Collaboration reported their measurements for $e^+e^- \to \pi^+ \pi^-$ up to 3 GeV. Recent analysis indicates there exists a resonance in the cross sections for $e^+e^- \to \pi^+\pi^-$ near 2.2 GeV with the resonance parameters to be, 
\begin{eqnarray}
m_Y= (2232 \pm 8 \pm 9 ) \ \mathrm{MeV}\nonumber\\
\Gamma_Y=(133\pm 14 \pm 4 )\ \mathrm{MeV}, \label{Eq:RP-BaBar}
\end{eqnarray}
which are well consistent with those of $X(2240)$~\cite{Ablikim:2018iyx}. 

Here, we adopt the same interference scenario to fit the cross sections for $e^+e^- \to \pi^+ \pi^-$, where the resonance parameters are considered as free parameters. The fitted parameters are collected in Table \ref{Tab:Para-pipi} as Mode I. From our fit, we find the mass of the vector resonance is similar to the one of BaBar fit as shown in Eq. (\ref{Eq:RP-BaBar}) but the width is much small, which is $\Gamma= (34.3\pm 72.1)$ MeV. The $\chi^2$ is estimated to be 2.11. In Mode I, the uncertainty of the vector meson resonance are particularly large and its center value are much different with the BaBar analysis. With the center values of the fitted parameters, we can obtain the fitted curve, which is present in Fig.~\ref{Fig:Fit-pipi} and we find the fitted curve can well reproduce the experimental data. 

To check the effect of the resonance widths in our fit, we fixed the resonance parameters of the vector meson as those of the BaBar analysis and fit the cross sections for $e^+ e^- \to \pi^+ \pi^-$. The fitted parameters are present in Table \ref{Tab:Para-pipi} as Mode II and the $\chi^2$ is estimated to be 2.78, which is a bit larger than the one of Mode I. From the dashed curve in Fig.~\ref{Fig:Fit-pipi}, one can find Mode II can also reproduce the experimental data. Comparing to Mode I, one can find the fitted curve can only reach up the minimum of the experimental data at $\sqrt{s}$=2.275 GeV. From our fit, we find the $\chi^2/\mathrm{n.d.f}$ is much smaller than one due to the large uncertainties of the experimental data. 

To further check the relation of the structures in $e^+ e^- \to K^+ K^-$ and $e^+e^- \to \pi^+ \pi^-$, we further fixed the resonance parameters to be those of $Y(2175)$ as shown in Eq.~(\ref{Eq:PDG-Y2175}). The fitted parameter values are presented in Table~\ref{Tab:Para-pipi} as Mode III and the $\chi^2$ is estimated to be 4.55, which is a bit larger than those of Mode I and Mode II. However, the $\chi^2/\mathrm{n.d.f}$ is 4.55/6, which indicates such a fit is also acceptable. The fitted curve is also  present in Fig.~\ref{Fig:Fit-pipi} for comparison. From the figure, one can find $Y(2175)$ can also produce a structure near 2.2 GeV, which is similar to the one of Mode II.

In the present interference scenario, we find the cross sections for $e^+ e^- \to \pi^+ \pi^-$ can also be reproduced with an acceptable $\chi^2$, but the resonance parameter can be different with different fitting assumptions. The resonance $Y(2175)$ can also produce a structure near 2.2 GeV. Due to the large uncertainties of the experimental data, we cannot draw any solid conclusion form the present fit. We expect the BES III and Belle II Collaborations can provide more precise data for this process in the future, which can help us to reveal the source of the structure near 2.2 GeV in the cross sections for $e^+e^- \to \pi^+ \pi^-$.

\section{Summary}
\label{Sec:Sum}

The strangeonium-like state $Y(2175)$ has been observed in various hidden strange channels. Its nature is a long-standing puzzle. There are some different interpretations to the inner structure of $Y(2175)$,  such as tetraquark, molecular, hybrid and conventional strangeonium. The open strange decays are important criteria of different interpretations. Thus, searching $Y(2175)$ in the open strange channels are crucial to understanding the nature of $Y(2175)$. 

In 2018, the BES III Collaboration reported their precise measurements of the cross sections for $e^+e^- \to K^+K^-$. No evidence of $Y(2175)$ was found, but a new structure around 2240 MeV, named $X(2240)$, was observed. In the present work, we consider the interference between the direct coupling and the vector resonance intermediate contributions to fit the cross sections for $e^+ e^- \to K^+ K^-$. We find the structure around 2240 MeV is resulted from $Y(2175)$ and the interference between two mechanisms are very important, which shifts the peak of $Y(2175)$ to 2240 MeV.

In our fit, we find that there are only two almost degenerate data from 2.31 to 2.5 GeV. The experimental data in this center-of-mass energy rang are important to determine the lineshape of the cross sections. We expect the BES III Collaboration could provide more precise data in this energy rang, which can help us to better understand the cross sections for $e^+e^- \to K^+ K^-$. 

We also extend the present interference scenario to investigate the cross sections for $e^+e^- \to \pi^+ \pi^-$, where a structure near 2.2 GeV were also reported by BaBar Collaboration. We find that $Y(2175)$ can also produce a structure near 2.2 GeV, but we cannot draw any solid conclusion due to the large uncertainties of the experimental data.

\section{ACKNOWLEDGMENTS} D.C. would like to thank Dr. 
Wen-Biao Yan and Bei-Jiang Liu for useful discussions of the BES data. This work is supported by the National Natural Science Foundation of China under the Grant Nos. 11775050 and 11675228.


\begin{thebibliography}{99}

\bibitem{Klempt:2007cp} 
  E.~Klempt and A.~Zaitsev,
  Phys.\ Rept.\  {\bf 454}, 1 (2007)
  doi:10.1016/j.physrep.2007.07.006
  [arXiv:0708.4016 [hep-ph]].

\bibitem{Brambilla:2010cs}
  N.~Brambilla {\it et al.},
  Eur.\ Phys.\ J.\ C {\bf 71} (2011) 1534
  doi:10.1140/epjc/s10052-010-1534-9
  [arXiv:1010.5827 [hep-ph]].

\bibitem{Chen:2016qju}
  H.~X.~Chen, W.~Chen, X.~Liu and S.~L.~Zhu,
  Phys.\ Rept.\  {\bf 639} (2016) 1
  doi:10.1016/j.physrep.2016.05.004
  [arXiv:1601.02092 [hep-ph]].
  
\bibitem{Lebed:2016hpi}
  R.~F.~Lebed, R.~E.~Mitchell and E.~S.~Swanson,
  Prog.\ Part.\ Nucl.\ Phys.\  {\bf 93} (2017) 143
  doi:10.1016/j.ppnp.2016.11.003
  [arXiv:1610.04528 [hep-ph]].

\bibitem{Guo:2017jvc}
  F.~K.~Guo, C.~Hanhart, U.~G.~Meißner, Q.~Wang, Q.~Zhao and B.~S.~Zou,
  Rev.\ Mod.\ Phys.\  {\bf 90} (2018) no.1,  015004
  doi:10.1103/RevModPhys.90.015004
  [arXiv:1705.00141 [hep-ph]].
  
\bibitem{Esposito:2016noz}
  A.~Esposito, A.~Pilloni and A.~D.~Polosa,
  Phys.\ Rept.\  {\bf 668} (2017) 1
  doi:10.1016/j.physrep.2016.11.002
  [arXiv:1611.07920 [hep-ph]].
  
\bibitem{Ali:2017jda}
  A.~Ali, J.~S.~Lange and S.~Stone,
  Prog.\ Part.\ Nucl.\ Phys.\  {\bf 97} (2017) 123
  doi:10.1016/j.ppnp.2017.08.003
  [arXiv:1706.00610 [hep-ph]].
    
\bibitem{Liu:2019zoy}
  Y.~R.~Liu, H.~X.~Chen, W.~Chen, X.~Liu and S.~L.~Zhu,
  Prog.\ Part.\ Nucl.\ Phys.\  {\bf 107} (2019) 237
  doi:10.1016/j.ppnp.2019.04.003
  [arXiv:1903.11976 [hep-ph]].


\bibitem{Brambilla:2019esw}
  N.~Brambilla, S.~Eidelman, C.~Hanhart, A.~Nefediev, C.~P.~Shen, C.~E.~Thomas, A.~Vairo and C.~Z.~Yuan,
  arXiv:1907.07583 [hep-ex].
  
\bibitem{Dong:2017gaw}
  Y.~Dong, A.~Faessler and V.~E.~Lyubovitskij,
  Prog.\ Part.\ Nucl.\ Phys.\  {\bf 94} (2017) 282.
  doi:10.1016/j.ppnp.2017.01.002
  
\bibitem{Aubert:2006bu}
  B.~Aubert {\it et al.}  [BABAR Collaboration],
  Phys.\ Rev.\  D {\bf 74}, 091103 (2006)
  [arXiv:hep-ex/0610018].



\bibitem{Ablikim:2007ab} 
  M.~Ablikim {\it et al.} [BES Collaboration],
  Phys.\ Rev.\ Lett.\  {\bf 100}, 102003 (2008)
  doi:10.1103/PhysRevLett.100.102003
  [arXiv:0712.1143 [hep-ex]].
  
\bibitem{Shen:2009zze}
  C.~P.~Shen {\it et al.}  [Belle Collaboration],
  Phys.\ Rev.\  D {\bf 80}, 031101 (2009)
  [arXiv:0808.0006 [hep-ex]].
  
\bibitem{Lees:2011zi} 
  J.~P.~Lees {\it et al.} [BaBar Collaboration],
  Phys.\ Rev.\ D {\bf 86}, 012008 (2012)
  doi:10.1103/PhysRevD.86.012008
  [arXiv:1103.3001 [hep-ex]].

\bibitem{Aubert:2007ym} 
  B.~Aubert {\it et al.} [BaBar Collaboration],
  Phys.\ Rev.\ D {\bf 77}, 092002 (2008)
  doi:10.1103/PhysRevD.77.092002
  [arXiv:0710.4451 [hep-ex]].

\bibitem{Aubert:2007ur} 
  B.~Aubert {\it et al.} [BaBar Collaboration],
  Phys.\ Rev.\ D {\bf 76}, 012008 (2007)
  doi:10.1103/PhysRevD.76.012008
  [arXiv:0704.0630 [hep-ex]].

\bibitem{Ablikim:2014pfc} 
  M.~Ablikim {\it et al.} [BESIII Collaboration],
  Phys.\ Rev.\ D {\bf 91}, no. 5, 052017 (2015)
  doi:10.1103/PhysRevD.91.052017
  [arXiv:1412.5258 [hep-ex]].
  
  
\bibitem{Ablikim:2019tpp} 
  M.~Ablikim {\it et al.} [BESIII Collaboration],
  Phys.\ Rev.\ D {\bf 100}, no. 3, 032009 (2019)
  doi:10.1103/PhysRevD.100.032009
  [arXiv:1907.06015 [hep-ex]].



\bibitem{Ablikim:2020pgw} 
  M.~Ablikim {\it et al.} [BESIII Collaboration],
  Phys.\ Rev.\ Lett.\  {\bf 124}, no. 11, 112001 (2020)
  doi:10.1103/PhysRevLett.124.112001
  [arXiv:2001.04131 [hep-ex]].


\bibitem{Ablikim:2020coo} 
  M.~Ablikim {\it et al.} [BESIII Collaboration],
  arXiv:2003.13064 [hep-ex].
  
  
  
  
\bibitem{Tanabashi:2018oca}
  M.~Tanabashi {\it et al.} [Particle Data Group],
  Phys.\ Rev.\ D {\bf 98} (2018) no.3,  030001.
  doi:10.1103/PhysRevD.98.030001
   	
\bibitem{Ablikim:2018iyx} 
  M.~Ablikim {\it et al.} [BESIII Collaboration],
  Phys.\ Rev.\ D {\bf 99}, no. 3, 032001 (2019)
  doi:10.1103/PhysRevD.99.032001
  [arXiv:1811.08742 [hep-ex]].

\bibitem{Aubert:2005rm} 
  B.~Aubert {\it et al.} [BaBar Collaboration],
  Phys.\ Rev.\ Lett.\  {\bf 95}, 142001 (2005)
  doi:10.1103/PhysRevLett.95.142001
  [hep-ex/0506081].

\bibitem{Aubert:2007zz} 
  B.~Aubert {\it et al.} [BaBar Collaboration],
  Phys.\ Rev.\ Lett.\  {\bf 98}, 212001 (2007)
  doi:10.1103/PhysRevLett.98.212001
  [hep-ex/0610057].

\bibitem{Abe:2007tk} 
  K.~F.~Chen {\it et al.} [Belle Collaboration],
  Phys.\ Rev.\ Lett.\  {\bf 100}, 112001 (2008)
  doi:10.1103/PhysRevLett.100.112001
  [arXiv:0710.2577 [hep-ex]].

\bibitem{Wang:2006ri} 
  Z.~G.~Wang,
  Nucl.\ Phys.\ A {\bf 791}, 106 (2007)
  doi:10.1016/j.nuclphysa.2007.04.012
  [hep-ph/0610171].

\bibitem{Drenska:2008gr} 
  N.~V.~Drenska, R.~Faccini and A.~D.~Polosa,
  Phys.\ Lett.\ B {\bf 669}, 160 (2008)
  doi:10.1016/j.physletb.2008.09.038
  [arXiv:0807.0593 [hep-ph]].

\bibitem{Deng:2010zzd} 
  C.~Deng, J.~Ping, F.~Wang and T.~Goldman,
  Phys.\ Rev.\ D {\bf 82}, 074001 (2010).
  doi:10.1103/PhysRevD.82.074001

\bibitem{Ding:2006ya} 
  G.~J.~Ding and M.~L.~Yan,
  Phys.\ Lett.\ B {\bf 650}, 390 (2007)
  doi:10.1016/j.physletb.2007.05.026
  [hep-ph/0611319].

\bibitem{Ding:2007pc} 
  G.~J.~Ding and M.~L.~Yan,
  Phys.\ Lett.\ B {\bf 657}, 49 (2007)
  doi:10.1016/j.physletb.2007.10.020
  [hep-ph/0701047].

\bibitem{Dudek:2011bn} 
  J.~J.~Dudek,
  Phys.\ Rev.\ D {\bf 84}, 074023 (2011)
  doi:10.1103/PhysRevD.84.074023
  [arXiv:1106.5515 [hep-ph]].

\bibitem{Ho:2019org} 
  J.~Ho, R.~Berg, T.~G.~Steele, W.~Chen and D.~Harnett,
  Phys.\ Rev.\ D {\bf 100}, no. 3, 034012 (2019)
  doi:10.1103/PhysRevD.100.034012
  [arXiv:1905.12779 [hep-ph]].

\bibitem{MartinezTorres:2008gy}
  A.~Martinez Torres, K.~P.~Khemchandani, L.~S.~Geng, M.~Napsuciale and E.~Oset,
  Phys.\ Rev.\ D {\bf 78} (2008) 074031
  doi:10.1103/PhysRevD.78.074031
  [arXiv:0801.3635 [nucl-th]].

\bibitem{Barnes:2002mu} 
  T.~Barnes, N.~Black and P.~R.~Page,
  Phys.\ Rev.\ D {\bf 68}, 054014 (2003)
  doi:10.1103/PhysRevD.68.054014
  [nucl-th/0208072].

\bibitem{Wang:2012wa}
  X.~Wang, Z.~F.~Sun, D.~Y.~Chen, X.~Liu and T.~Matsuki,
  Phys.\ Rev.\ D {\bf 85} (2012) 074024
  doi:10.1103/PhysRevD.85.074024
  [arXiv:1202.4139 [hep-ph]].

\bibitem{Azizi:2019ecm} 
  K.~Azizi, S.~S.~Agaev and H.~Sundu,
  Nucl.\ Phys.\ B {\bf 948}, 114789 (2019)
  doi:10.1016/j.nuclphysb.2019.114789
  [arXiv:1906.04061 [hep-ph]].


\bibitem{Lu:2019ira}
  Q.~F.~Lü, K.~L.~Wang and Y.~B.~Dong,
  arXiv:1903.05007 [hep-ph].

\bibitem{Zhu:2019ibc}
  J.~T.~Zhu, Y.~Liu, D.~Y.~Chen, L.~Jiang and J.~He,
  arXiv:1911.03706 [hep-ph].

\bibitem{Wang:2019jch} 
  Y.~r.~Wang, J.~f.~Hu, C.~Q.~Pang and T.~J.~Zhang,
  arXiv:1910.12408 [hep-ph].
  
\bibitem{Chen:2010nv} 
  D.~Y.~Chen, J.~He and X.~Liu,
  Phys.\ Rev.\ D {\bf 83}, 054021 (2011)
  doi:10.1103/PhysRevD.83.054021
  [arXiv:1012.5362 [hep-ph]].

\bibitem{Chen:2011kc} 
  D.~Y.~Chen, J.~He and X.~Liu,
  Phys.\ Rev.\ D {\bf 83}, 074012 (2011)
  doi:10.1103/PhysRevD.83.074012
  [arXiv:1101.2474 [hep-ph]].

\bibitem{Chen:2015bft}
  D.~Y.~Chen, X.~Liu, X.~Q.~Li and H.~W.~Ke,
  Phys.\ Rev.\ D {\bf 93} (2016) 014011
  doi:10.1103/PhysRevD.93.014011
  [arXiv:1512.04157 [hep-ph]].
  
\bibitem{Chen:2017uof}
  D.~Y.~Chen, X.~Liu and T.~Matsuki,
  Eur.\ Phys.\ J.\ C {\bf 78} (2018) no.2,  136
  doi:10.1140/epjc/s10052-018-5635-1
  [arXiv:1708.01954 [hep-ph]].


\bibitem{Aubert:2007ef}
  B.~Aubert {\it et al.} [BaBar Collaboration],
  Phys.\ Rev.\ D {\bf 76} (2007) 092005
   Erratum: [Phys.\ Rev.\ D {\bf 77} (2008) 119902]
  doi:10.1103/PhysRevD.77.119902, 10.1103/PhysRevD.76.092005
  [arXiv:0708.2461 [hep-ex]].


\bibitem{Lees:2012cj} 
  J.~P.~Lees {\it et al.} [BaBar Collaboration],
  Phys.\ Rev.\ D {\bf 86}, 032013 (2012)
  doi:10.1103/PhysRevD.86.032013
  [arXiv:1205.2228 [hep-ex]].

\bibitem{Chen:2011cj} 
  D.~Y.~Chen, X.~Liu and T.~Matsuki,
  Eur.\ Phys.\ J.\ C {\bf 72}, 2008 (2012)
  doi:10.1140/epjc/s10052-012-2008-z
  [arXiv:1112.3773 [hep-ph]].

\bibitem{Bauer:1977iq} 
  T.~H.~Bauer, R.~D.~Spital, D.~R.~Yennie and F.~M.~Pipkin,
  Rev.\ Mod.\ Phys.\  {\bf 50}, 261 (1978)
  Erratum: [Rev.\ Mod.\ Phys.\  {\bf 51}, 407 (1979)].
  doi:10.1103/RevModPhys.50.261

\bibitem{Bauer:1975bv} 
  T.~Bauer and D.~R.~Yennie,
  Phys.\ Lett.\  {\bf 60B}, 165 (1976).
  doi:10.1016/0370-2693(76)90414-7

\bibitem{Bauer:1975bw} 
  T.~Bauer and D.~R.~Yennie,
  Phys.\ Lett.\  {\bf 60B}, 169 (1976).
  doi:10.1016/0370-2693(76)90415-9

\bibitem{Yang:2019mzq}
  Y.~Yang, D.~Y.~Chen and Z.~Lu,
  Phys.\ Rev.\ D {\bf 100} (2019) no.7,  073007
  doi:10.1103/PhysRevD.100.073007
  [arXiv:1902.01242 [hep-ph]].

\bibitem{Chen:2008hka}
  D.~Y.~Chen, H.~Q.~Zhou and Y.~B.~Dong,
  Phys.\ Rev.\ C {\bf 78} (2008) 045208
  doi:10.1103/PhysRevC.78.045208
  [arXiv:0806.2489 [nucl-th]].
  
\bibitem{Lees:2013gzt} 
  J.~P.~Lees {\it et al.} [BaBar Collaboration],
  Phys.\ Rev.\ D {\bf 88}, no. 3, 032013 (2013)
  doi:10.1103/PhysRevD.88.032013
  [arXiv:1306.3600 [hep-ex]].
  
    
\bibitem{Shen:2009mr}
  C.~P.~Shen and C.~Z.~Yuan,
  Chin.\ Phys.\ C {\bf 34} (2010) 1045
  doi:10.1088/1674-1137/34/8/002
  [arXiv:0911.1591 [hep-ex]].

\bibitem{BABAR:2019oes}
  J.~P.~Lees {\it et al.} [BaBar Collaboration],
  Phys.\ Rev.\ D {\bf 101} (2020) no.1,  012011
  doi:10.1103/PhysRevD.101.012011
  [arXiv:1912.04512 [hep-ex]].
  
\bibitem{Lees:2012cj} 
  J.~P.~Lees {\it et al.} [BaBar Collaboration],
  Phys.\ Rev.\ D {\bf 86}, 032013 (2012)
  doi:10.1103/PhysRevD.86.032013
  [arXiv:1205.2228 [hep-ex]].
  
\end{thebibliography}
\end{document}